\documentclass[%
 aip,
 amsmath,amssymb,
 reprint,%
]{revtex4-1}

\usepackage{graphicx}
\usepackage{dcolumn}
\usepackage{bm}
\usepackage{upgreek}
\usepackage[utf8]{inputenc}
\usepackage[T1]{fontenc}
\usepackage{mathptmx}
\usepackage{subfigure}
\usepackage{amsmath}
\usepackage{amsfonts}
\usepackage{mathtools}
\usepackage{wasysym}
\usepackage{amssymb}
\usepackage{subfigure}
\usepackage{ragged2e}
\usepackage{upgreek}
\usepackage{graphicx}

\begin{document}

\preprint{AIP/123-QED}

\title[Frequency Shift Algorithm: Application to a FDM Readout of X-ray TES Microcalorimeters]{Frequency Shift Algorithm: Application to a Frequency-Domain Multiplexing Readout of X-ray Transition-Edge Sensor Microcalorimeters}

\author{D.~Vaccaro}
 \email{d.vaccaro@sron.nl}
\affiliation{NWO-I/SRON Netherlands Institute for Space Research, Sorbonnelaan 2, 3584 CA Utrecht, The Netherlands}
\author{H.~Akamatsu}
\affiliation{NWO-I/SRON Netherlands Institute for Space Research, Sorbonnelaan 2, 3584 CA Utrecht, The Netherlands}
\author{J.~van~der~Kuur}
\affiliation{NWO-I/SRON Netherlands Institute for Space Research, Landleven 12, 9747 AD Groningen, The Netherlands}
\author{P.~van~der~Hulst}
\affiliation{NWO-I/SRON Netherlands Institute for Space Research, Sorbonnelaan 2, 3584 CA Utrecht, The Netherlands}
\author{A.C.T.~Nieuwenhuizen}
\affiliation{NWO-I/SRON Netherlands Institute for Space Research, Sorbonnelaan 2, 3584 CA Utrecht, The Netherlands}
\author{P.~van~Winden}
\affiliation{NWO-I/SRON Netherlands Institute for Space Research, Sorbonnelaan 2, 3584 CA Utrecht, The Netherlands}
\author{L.~Gottardi}
\affiliation{NWO-I/SRON Netherlands Institute for Space Research, Sorbonnelaan 2, 3584 CA Utrecht, The Netherlands}
\author{R.~den~Hartog}
\affiliation{NWO-I/SRON Netherlands Institute for Space Research, Landleven 12, 9747 AD Groningen, The Netherlands}
\author{M.P.~Bruijn}
\affiliation{NWO-I/SRON Netherlands Institute for Space Research, Sorbonnelaan 2, 3584 CA Utrecht, The Netherlands}
\author{M.~D'Andrea}
\affiliation{NWO-I/SRON Netherlands Institute for Space Research, Sorbonnelaan 2, 3584 CA Utrecht, The Netherlands}
\author{J.R.~Gao}
\affiliation{NWO-I/SRON Netherlands Institute for Space Research, Sorbonnelaan 2, 3584 CA Utrecht, The Netherlands}
\affiliation{Optics Group, Department of Imaging Physics, Delft University of Technology, Delft, 2628 CJ, the Netherlands}
\author{J.W.A.~den~Herder}
\affiliation{NWO-I/SRON Netherlands Institute for Space Research, Sorbonnelaan 2, 3584 CA Utrecht, The Netherlands}
\author{R.W.M. Hoogeveen}
\affiliation{NWO-I/SRON Netherlands Institute for Space Research, Sorbonnelaan 2, 3584 CA Utrecht, The Netherlands}
\author{B.~Jackson}
\affiliation{NWO-I/SRON Netherlands Institute for Space Research, Landleven 12, 9747 AD Groningen, The Netherlands}
\author{A.J.~van~der~Linden}
\affiliation{NWO-I/SRON Netherlands Institute for Space Research, Sorbonnelaan 2, 3584 CA Utrecht, The Netherlands}
\author{K.~Nagayoshi}
\affiliation{NWO-I/SRON Netherlands Institute for Space Research, Sorbonnelaan 2, 3584 CA Utrecht, The Netherlands}
\author{K.~Ravensberg}
\affiliation{NWO-I/SRON Netherlands Institute for Space Research, Sorbonnelaan 2, 3584 CA Utrecht, The Netherlands}
\author{M.L.~Ridder}
\affiliation{NWO-I/SRON Netherlands Institute for Space Research, Sorbonnelaan 2, 3584 CA Utrecht, The Netherlands}
\author{E.~Taralli}
\affiliation{NWO-I/SRON Netherlands Institute for Space Research, Sorbonnelaan 2, 3584 CA Utrecht, The Netherlands}
\author{M.~de~Wit}
\affiliation{NWO-I/SRON Netherlands Institute for Space Research, Sorbonnelaan 2, 3584 CA Utrecht, The Netherlands}

\date{\today}

\begin{abstract}
In the frequency-domain multiplexing (FDM) scheme, transition-edge sensors (TES) are individually coupled to superconducting LC filters and AC biased at MHz frequencies through a common readout line. To make efficient use of the available readout bandwidth and to minimize the effect of non-linearities, the LC resonators are usually designed to be on a regular grid. The lithographic processes however pose a limit on the accuracy of the effective filter resonance frequencies. Off-resonance bias carriers could be used to suppress the impact of intermodulation distortions, which nonetheless would significantly affect the effective bias circuit and the detector spectral performance. In this paper we present a frequency shift algorithm (FSA) to allow off-resonance readout of TES's while preserving the on-resonance bias circuit and spectral performance, demonstrating its application to the FDM readout of a X-ray TES microcalorimeter array. We discuss the benefits in terms of mitigation of the impact of intermodulation distortions at the cost of increased bias voltage and the scalability of the algorithm to multi-pixel FDM readout. We show that with FSA, in multi-pixel and frequencies shifted on-grid, the line noises due to intermodulation distortion are placed away from the sensitive region in the TES response and the X-ray performance is consistent with the single-pixel, on-resonance level.
\end{abstract}

\maketitle

\begin{quotation}
The following paper has been accepted for publication in \textit{Review of Scientific Instruments}.
\end{quotation}

\section{Introduction}

Transition-edge sensors \cite{tes} (TES) are employed as very sensitive photon detectors in a wide number of experimental applications, either ground-based, balloon-borne or satellite telescopes, as bolometers or microcalorimeters for mm-wave radiation up to gamma rays, such as LiteBIRD\cite{litebird}, SPICA-SAFARI\cite{safari}, Athena X-IFU\cite{athena}, LSPE-SWIPE\cite{lspe}, QUBIC\cite{qubic}.

To comply with the stringent thermal requirements to operate large arrays of TES's at sub-Kelvin temperatures and to reduce wiring complexity and cost, their readout is performed using a multiplexing scheme, in which the signals of the detectors are modulated with a set of orthogonal carriers in a limited information bandwidth through a common transmission line. The sum of the signals is fed to a cold superconducting quantum interference device (SQUID) amplifier stage, then the output is retrieved back to warm, digital electronics, where demodulation of the individual TES signals is performed.

\begin{figure*}
\includegraphics[width=16cm]{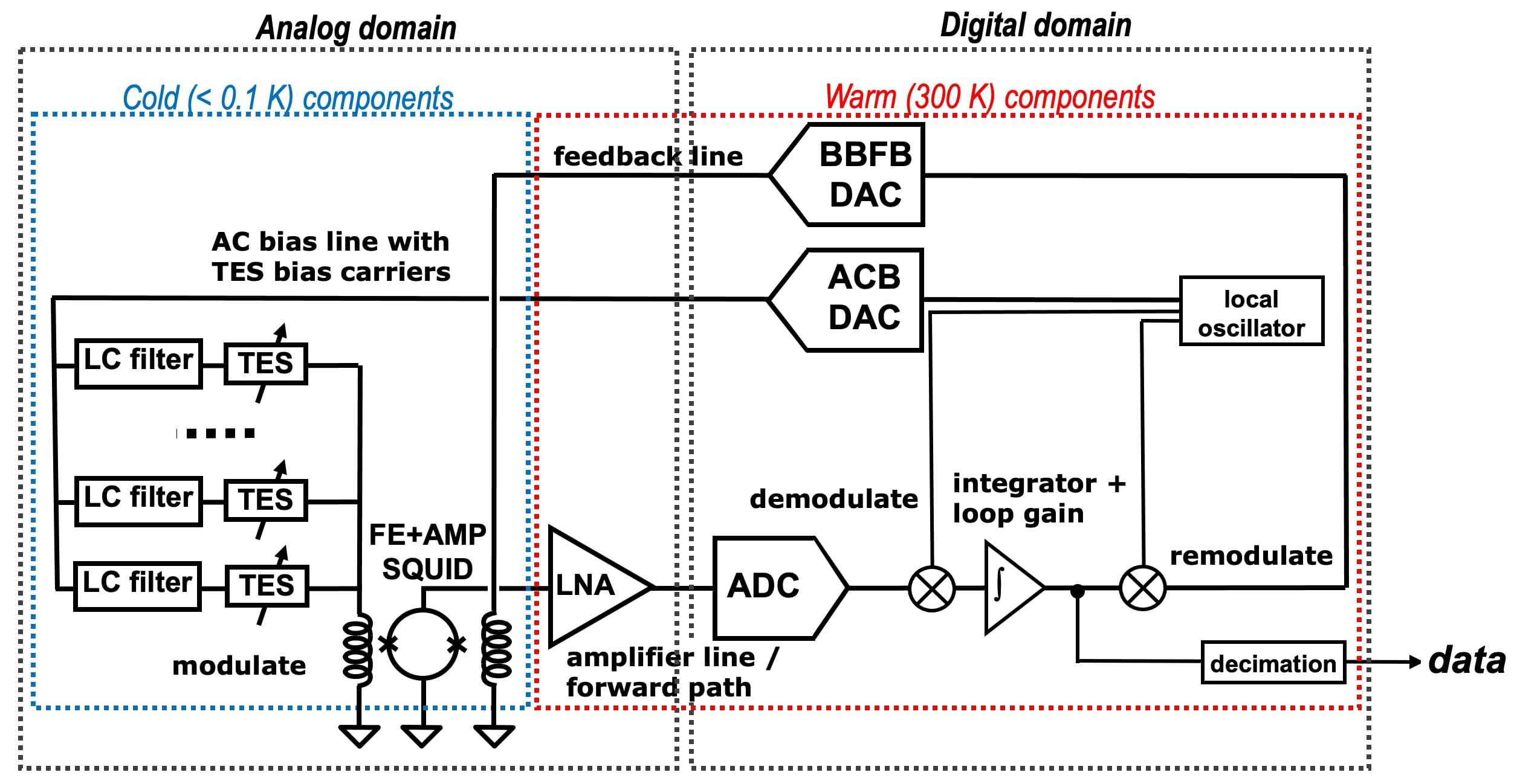}
\caption{Diagram showing our FDM readout scheme with base-band feedback. AC bias carriers are generated by the FPGA and converted to analog via the AC bias DAC from our custom digital electronics (DEMUX) board at room temperature. SQUIDs are controlled via a custom analog front-end electronics (FEE). The SQUID output signal is first pre-amplified via a low noise amplifier in the FEE, then fed via an ADC to the DEMUX board, where demodulation is performed to retrieve the individual TES's signals. These signals are re-converted to analog via the Feedback DAC and used to null the error signal at the SQUID summing point just around each bias carrier. Such base-band feedback scheme allows an increase of the available dynamic range, so that a large number of pixels can be accommodated in one readout chain.}\label{fdm_scheme}
\end{figure*}

To readout arrays of TES-based X-ray microcalorimeters, we have been developing a frequency-domain multiplexing (FDM) readout technology with base-band feedback \cite{bbfb} (BBFB). In the FDM scheme (see Figure~\ref{fdm_scheme}), the detectors are voltage biased with sinusoidal carriers at different frequencies in the MHz range. A tuned, high-Q LC bandpass filter is put in series with each detector, with the double function of limiting the information bandwidth and allowing only one carrier to provide the AC bias. The TES's signals are then summed at the input of a SQUID and demodulated at room temperature by digital electronics.

The condition of independence between the carriers used to codify each pixel's information sets a minimum ratio between the carrier's frequency spacing and the individual pixel bandwidth, which in turn is set by the detector speed and stability conditions under negative electro-thermal feedback \cite{irwin_tau}. To make use of the available readout bandwidth in an efficient way, the LC filters are designed to fall on a regular grid, so that the bias carriers are integer multiples of a frequency spacing that is constant between each subsequent carrier. Due to unavoidable fabrication tolerances and stray inductance from wire bonding, in practice resonators do not match with the expected frequency, with a typical spread of $\approx$~3~kHz around the design value \cite{bruijn}.

Non-linearities in the readout chain, due $e.g.$ to SQUID amplifiers or DACs, can generate spurious signals appearing as line noises in the carrier sidebands. With a high number of multiplexed detectors, such intermodulation line noises (IMLN) are likely to fall within the sensitive region of the detector response, degrading the detector performance. The frequencies of IMLN's are defined by linear combinations of the bias frequencies, hence choosing the latter on a regular grid, so that also the IMLNs' frequencies would conveniently be on a regular grid outside of the TES sensitive band, would avoid such performance degradation. Given the high Q of the resonators however, off-resonance operation effectively changes the bias circuit, since the reactive impedance of the LC filter becomes no longer negligible compared to the TES resistance.

In a previous work \cite{hiroki2018} we showed that the spectral performance of one X-ray TES microcalorimeter, spoiled by IMLN's, could be recovered with off-resonance operation by employing an early version of a frequency shift algorithm (FSA). The original concept of FSA\cite{jan2018} contemplated the tuning of the LC resonance frequency by adding an active impedance in TES bias circuit via a feedback on the modulated bias signal. That version had the drawback of requiring parameter fine tuning for each pixel and frequency shift, thus not being scalable. In this paper we report on a new implementation of FSA, designed to allow scalability to multi-pixel operation. The technical aspects of such algorithm are discussed in a companion paper \cite{paulpaper}. The operation scheme is outlined at the beginning of Section~\ref{requirements}.

The paper is structured as follows. We first describe our experimental setup in Section~\ref{setup}, then in Section~\ref{requirements} the FSA working principle and requirements for the application on an X-ray microcalorimeter readout are discussed. In Section~\ref{iv_elec} the reproducibility of the bias circuit in off-resonance readout is tested and finally in Section~\ref{results} the results of X-ray experiments with FSA in single-pixel and multi-pixel configurations are reported.

\section{Experimental setup}\label{setup}

\begin{figure}[!h]
\centering
\includegraphics[width=8.5cm]{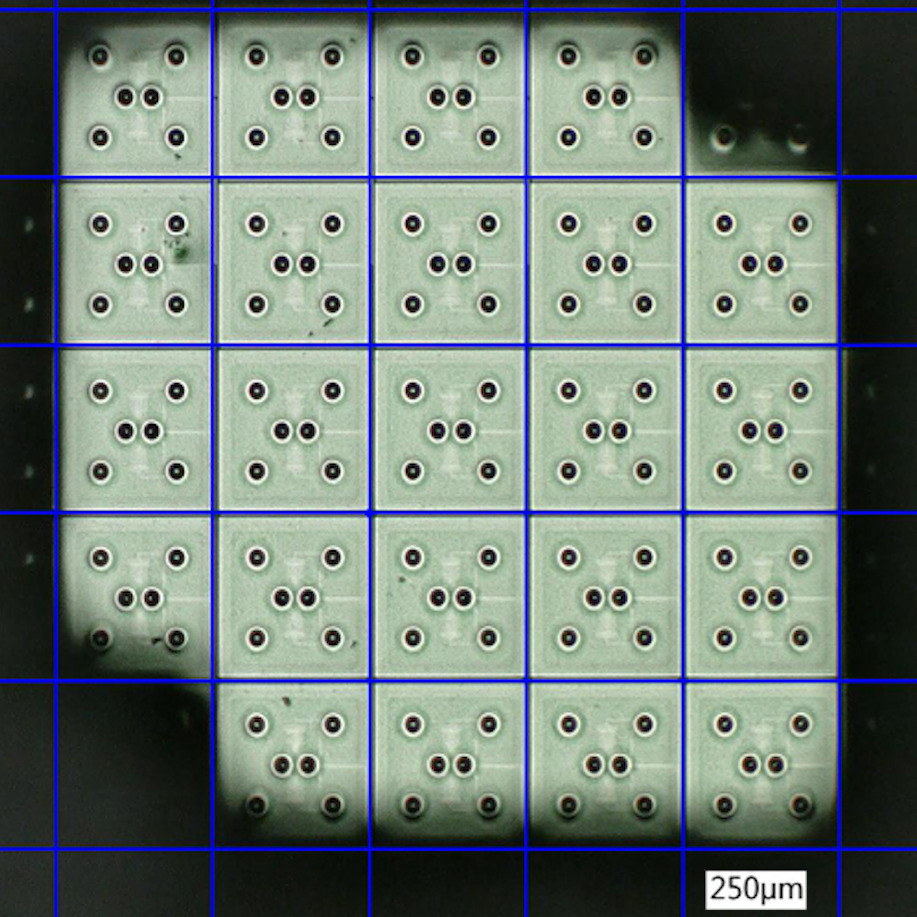}
\caption{Zoomed view through the Cu collimator of the SRON R1a 5$\times$5 TES array, mounted on the Cu holder. Top right and bottom left pixels are not connected, so that a total of 23 pixels is available.}\label{tes}
\end{figure}

For our experiments we used a 5$\times$5 TES microcalorimeter array developed at SRON Utrecht (for details about the fabrication, see K.~Nagayoshi~\emph{et al.}\cite{ken}). Each TES consists of a 100$\times$30 $\upmu$m$^{2}$ Ti/Au bilayer deposited on a SiN membrane and coupled to a 240$\times$240$~\upmu$m$^{2}$ gold absorber via two central pillars and four corner stems, as visible in Figure~\ref{tes}. These devices have $T_{C}~\simeq$~72~mK and R$_{N} \simeq$~75~m$\Omega$.

\begin{figure}[!h]
\centering
\includegraphics[width=8.5cm]{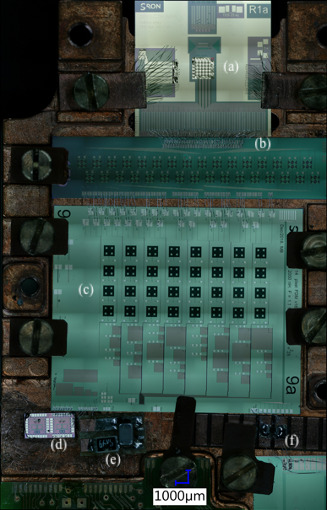}
\caption{Cryogenic components of our ``XFDM" setup mounted on the Cu sample holder: (a)~SRON TES array, (b) transformer chip, (c) LC filter chip, (d) VTT 6-series array front-end SQUID, (e) RC snubber in parallel to SQUID input coil and (f) SMD shunt resistor.}\label{xfdm}
\end{figure}

To perform the FDM readout the TES's are coupled to custom superconducting LC filters and transformers, to achieve impedance matching between the TES and LC filter. For this setup, LC filter and transformer chips result in a 2.5~$\upmu$H effective inductance. The detectors are voltage-biased through a 750~m$\Omega$ resistor and a capacitive divider with 1:26 ratio, for an effective shunt resistance of 1.4~m$\Omega$. Remanent magnetic field is cancelled by means of superconducting Helmholz coils. The cryogenic components are mounted on a oxygen-free high-conductivity (OFHC) copper holder, as depicted in Figure \ref{xfdm}, enclosed in a Nb shell and an outer Cryoperm shield. A $^{55}$Fe source is hosted on the superconducting magnetic shield (Nb shell) to exploit the Mn-K$_{\upalpha}$ line complex for measuring the X-ray spectral performance of the detectors at 5.9 keV, with a count rate of $\approx$ 1.5 counts per second.

The setup is cooled to cryogenic temperature by means of a Leiden Cryogenics dilution unit with 400~$\upmu$W of cooling power at 110 mK. For the measurements reported in this paper, the base temperature is 60 mK. The temperature of the setup is monitored via a Germanium thermistor anchored to the Cu holder. Mechanical vibrations are damped by suspending the setup to the mixing chamber with Kevlar wires. Thermal anchoring is performed by means of OFHC Cu braids. A 2-stage VTT SQUID (Model J3 as Front-End SQUID and model F5 as Amplifier SQUID) is employed as trans-impedance amplifier to pre-amplify the TES signals at cryogenic stage, which are demodulated and re-modulated for base-band feedback by an FPGA board, at a sampling rate of 40 MSPS. An RC snubber is placed in parallel to the SQUID input coil to lower the Q-factor of the 10's of MHz resonance created by the parallel between the SQUID input inductance and parasitic capacitance.

Temperature drifts in the cooling system sensed by the detectors are corrected in the data analysis pipeline exploiting the TES baseline current and the pulse height information. The energy non-linearity is calibrated out by using the zero energy information, the Mn-K$_{\upalpha}$ and Mn-K$_{\upbeta}$ (6.5 keV) lines. The energy of the X-ray event is evaluated by means of the optimal filtering technique. The spectral performance is then assessed by fitting the Mn-K$_{\upalpha}$ model \cite{holzer} to the collected events by using the Cash-statistics \cite{cstat} in the maximum-likelihood method. The collected events are typically 3-4 thousand in order to obtain sufficient statistics, with a typical statistical error on the energy resolution of 0.2~eV.

Considering the typical energy resolution of the 23 connected pixels, measured at R/R$_{N}$ typically in the range  0.15 to 0.25 and frequencies distributed to use the full 1-5~MHz readout bandwidth, the average single-pixel X-ray spectral performance is at a level of 2.7~eV at 5.9 keV.

\section{FSA for X-ray TES microcalorimeters}\label{requirements}

In this paper we focus on the application of FSA to the readout of X-ray TES microcalorimeters. We point out however that the algorithm is designed to be implementable also in different FDM readouts, such as for TES bolometer arrays as the one of SPICA-SAFARI \cite{safari}.

\subsection{FSA working principle}

The basic idea of FSA is that each carrier is shifted from the resonator center frequency and the carrier amplitude is compensated for the signal transfer loss due to the off-resonance operation of the LC filter. Such compensation is achieved by means of a proportional-integral controller, implemented in the FPGA. Similarly to the BBFB scheme, the FSA controller works in base-band to avoid stability issues.

\begin{figure}[!h]
\centering
\includegraphics[width=9cm]{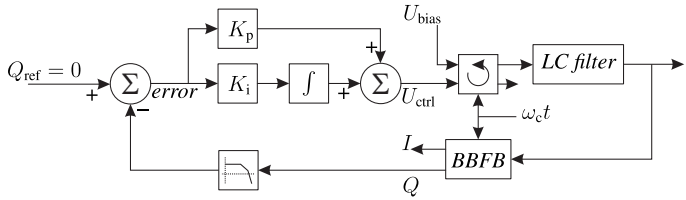}
\caption{Block diagram for the FSA controller \cite{paulpaper}. The quadrature BBFB output is the controller input, used as error signal with respect to the $Q=0$ reference. The controller loop is then split into two paths: an integrator with gain $K_{i}$ and a proportional controller with gain $K_{p}$. The ratio between the integrating and proportional transfer functions defines the crossover frequency $f_{PI}$. The complex vector given by the bias signal $U_{bias}$ (real) and the controller signal $U_{ctrl}$ (imaginary) is modulated (\emph{i.e.} rotated by an angle $\omega_{c}t$) and the real part of the resulting output (the AC carrier) is fed to the LC filter. The SQUID output is then demodulated, with only the $Q$-component of the BBFB output being used in the controller loop. To suppress the out-of-band $Q$-signal, a second order low-pass filter is employed.} \label{qnuller}
\end{figure}

In off-resonance operation, the imaginary impedance of the LC filter causes a phase difference between the TES current and bias voltage. The bias voltage is known and the TES current is measured via the BBFB, whose outputs are the demodulated in-phase ($I$) and quadrature ($Q$) orthogonal components of the TES current. Since the phase difference is correlated with the quadrature component, a convenient target for the controller is the $Q$ signal. The FSA controls the $Q$ output of the BBFB via a feedback scheme, explained in Figure \ref{qnuller}. In this "$Q$-nuller" scheme, the quadrature component of the TES current is the input of the controller, which uses it as an error signal with respect to a $Q=0$ reference.

For microcalorimeter application, a critical aspect is the speed of the controller, which should be sufficiently fast to compensate the large decrease in TES current when X-ray photons hit the detector. To make the controller faster, the integrator gain is increased, an action which in principle can make the controller unstable. To increase the stability margin of the controller a proportional path is implemented. The regime changes from integrating to proportional at frequencies higher than a crossover frequency $f_{PI}$, calculated by the ratio of the integrating and proportional transfer functions. In addition, a digital low-pass filter with cut-off frequency $f_{LPF}$ is present in the loop to prevent interference between the $Q$-feedback signals of neighbor pixels, fundamental for multi-pixel operation.

Substantially, the FSA allows to handle the LC filter spread by moving the issue from the cold components to the warm, digital readout electronics, which are more feasible to control the generation of intermodulation products, whose effect on performance is discussed as follows.

\subsection{Effect of intermodulation distortions}

The TES is sensitive to a IMLN when its frequency is close to the inverse of the effective time constant under electro-thermal feedback (thermal bandwidth). In this circumstance, the TES current is modulated and the energy signal depends on the phase at which the X-ray photon hits the detector, with the effect of severely degrading the measured energy resolution.

Ideally, it would be desirable to tune the carriers back to the original 100 kHz grid. Considering a 3$\upsigma$ scatter of the LC resonances around the design values, this would require a shift of the bias frequencies of the order of 10 kHz. A limitation of FSA is that the controller is expected to become unstable if the applied shift from resonance is larger than the resonator bandwidth. For the considered setup, the bandwidth of the RLC resonator with the TES biased at typical working point is of the order of 1 kHz. Given this boundary condition, shifting back the carriers to the 100 kHz grid is an unrealistic goal.

In practice, such extreme frequency shifts are not necessary. FDM for microcalorimeters requires TES's with ``slow" effective response time, so that a larger inductance can be employed to shrink down the electrical bandwidth $\approx$ R/L to prevent carrier leakage from neighbouring pixels~\cite{hiroki2018}, while ensuring stable TES operation. For this reason, our detectors are designed to have a thermal bandwidth of approximately 300~Hz, ``slow" enough to prevent carrier leakage at 100~kHz separation. This characteristic can be also exploited to effectively use frequency shifts much smaller than 10 kHz.

The idea is then to place the bias carriers on a regular grid, so that the IMLN's would fall exactly at the same frequency (and higher harmonics) in the TES response, sufficiently far from the thermal bandwidth. Therefore, we want to assess what is the nearest frequency grid (taking into account DAC resolution) for which no impact to the detector performance from intermodulation distortions is expected.

\begin{figure*}
\includegraphics[width=18cm]{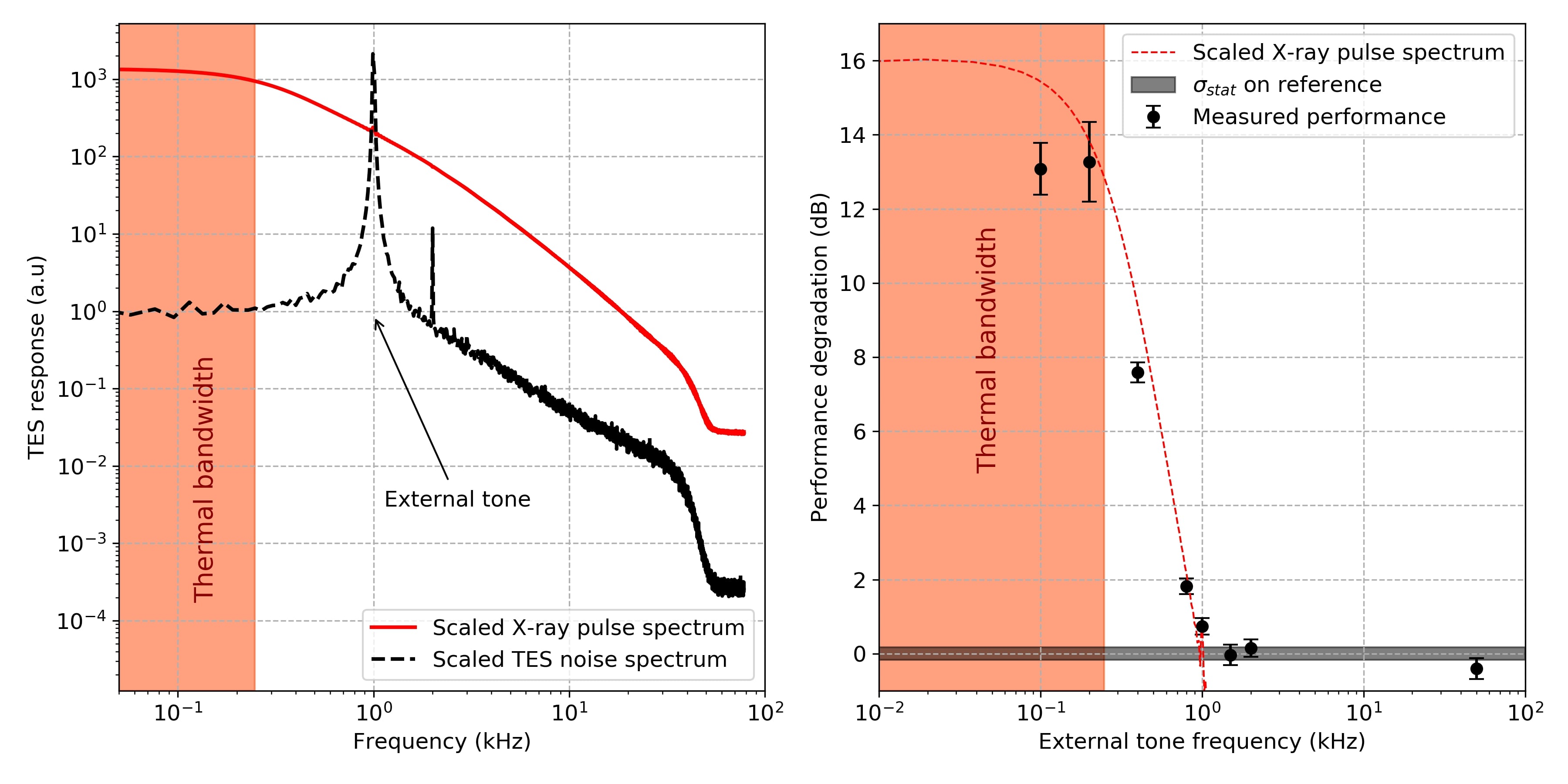}
\caption{\emph{Left panel}: X-ray pulse spectrum  and TES noise spectrum with applied external tone that mimics the effect of an IMLN. \emph{Right panel}: measured degradation of energy resolution with respect to the typical value for the considered pixel as a function of tone frequency $f_{tone}$, which is defined from the difference to the LC resonator frequency. The X-ray pulse spectrum is rescaled and overplotted in dashed red to show the thermal bandwidth, evaluated at the 3~dB point.}\label{imln_scan}
\end{figure*}

To do so, we measure the energy resolution of one pixel with an external tone applied to mimic the effect of an IMLN (Figure~\ref{imln_scan}, left panel). The amplitude of the tone is kept constant to a level of 3 orders of magnitude higher than the TES response. The frequency $f_{tone}$ is changed from values within the TES thermal bandwidth to the tail of the demodulated signal cut off by the decimation filter. In the right panel of Figure~\ref{imln_scan} the measured performance degradation with respect to the typical value for the investigated pixel is shown, scaled in dB units.

We observe that the X-ray performance severely deteriorates when the tone frequency falls into the thermal bandwidth and is recovered by shifting the tone frequency outside the thermal bandwidth, with no significant degradation for $f_{tone}=1.5$~kHz and a 0.6~dB degradation for $f_{tone}=1$~kHz, with respect to the typical energy resolution for the active pixel. From this experiment we conclude that, for such ``slow" devices, a 1.5~kHz grid would be necessary to completely tame out the impact of the applied external tone.

\begin{figure*}
\includegraphics[width=17cm]{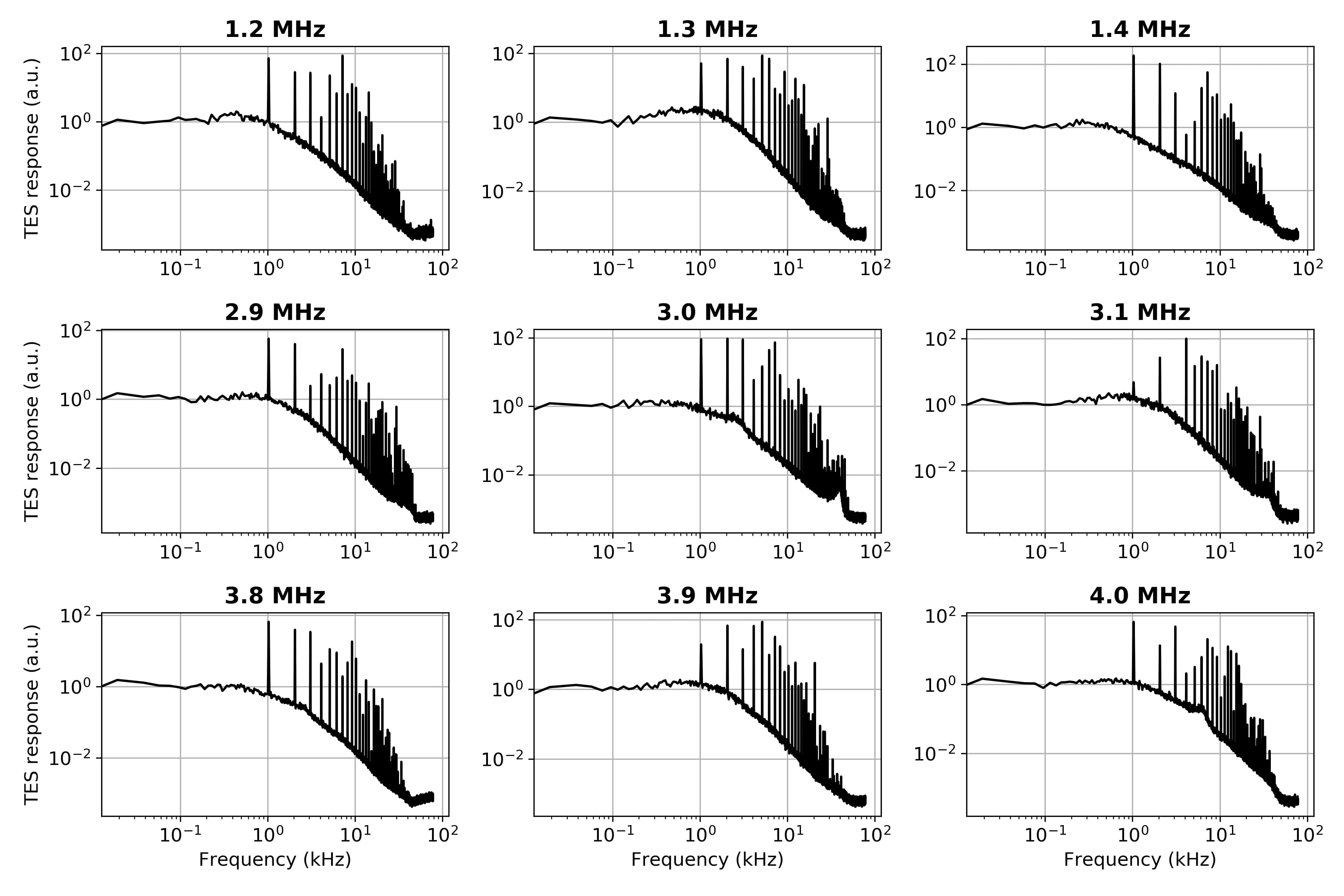}
\caption{TES noise spectra for 9 pixels (3 each for low-, mid- and high-frequency regime in 100~kHz separation) acquired in a 20-pixel multiplexing measurement. FSA is applied on all the pixels using a 1~kHz grid configuration. As a result, the bias frequencies are shifted such that all the IMLN's appear at integer multiples of 1~kHz.}\label{multispectra}
\end{figure*}

In Figure~\ref{multispectra} we show, as an example of a practical multiplexing case with a representative number of 20 pixels active, the noise spectra for 9 pixels in groups of 3 neighbours in 100~kHz separation in different bias frequency regimes. FSA is active on all the pixels and frequency shift is applied to place the carriers on a 1~kHz grid. As can be seen, the amplitude for the observed IMLN's is at least one order of magnitude lower compared to the applied tone in our experiment. Therefore, we expect the 0.6~dB degradation at $f_{tone}=1$~kHz to become negligible in a practical multi-pixel configuration. To minimize the frequency shifts necessary to arrange the carriers on grid, thus reducing the risk of shifting the carriers outside of the stability margin of the controller, we decide to perform our further tests in the 1~kHz grid configuration. Such choice also minimizes the impact on the bias voltage fed by the DAC, discussed in the next subsection.

\subsection{Impact on AC bias DAC}

The TES current is maintained to its on-resonance value when changing the bias frequency at the expense of an increase in the voltage fed by the bias DAC. A requirement for practical application is that such increase in DAC output should be acceptable from the point of view of (a) dynamic range and (b) noise contribution to the energy resolution.

\begin{figure*}
\subfigure{\includegraphics[width=8.5cm]{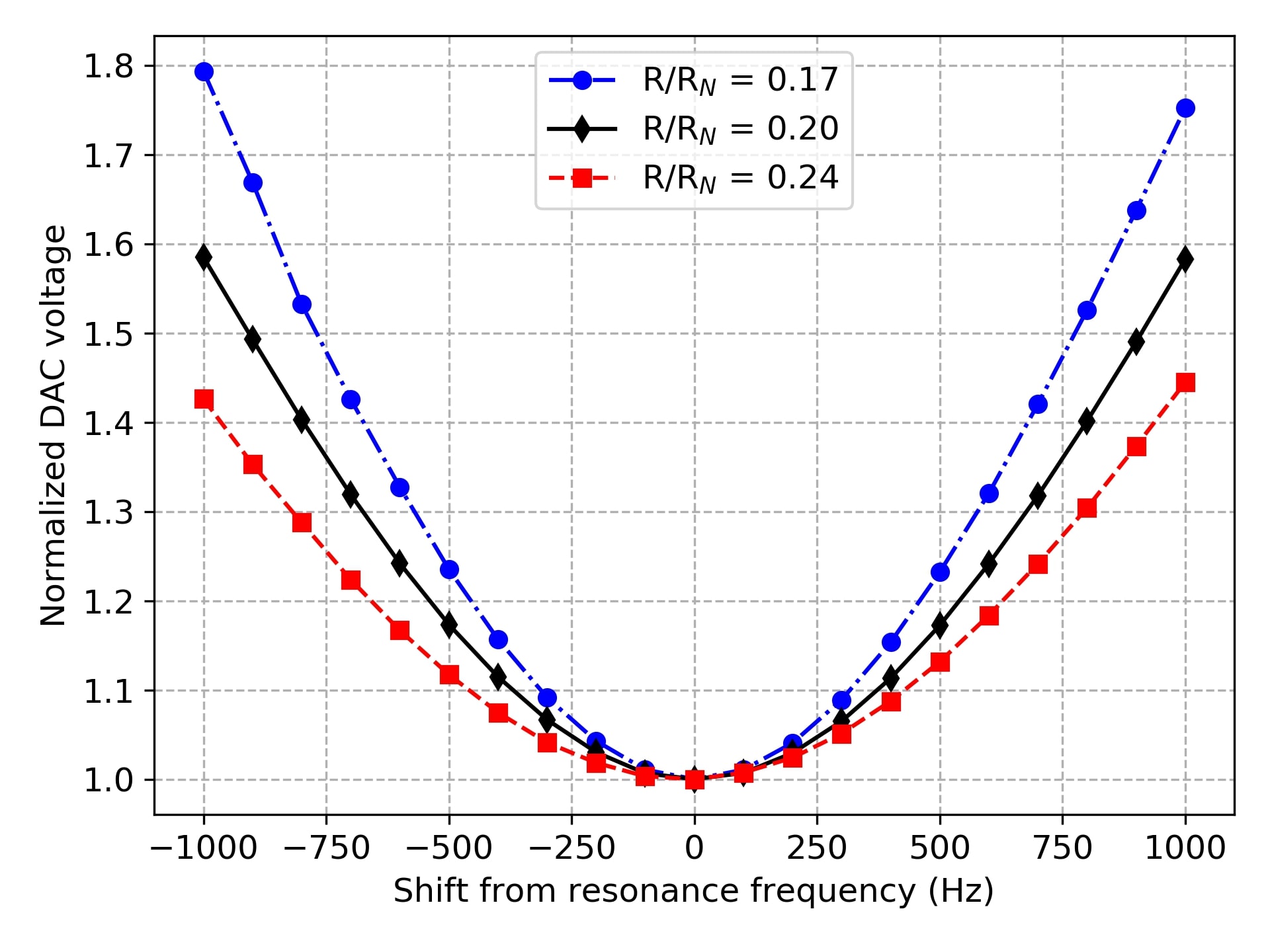}}
\subfigure{\includegraphics[width=8.5cm]{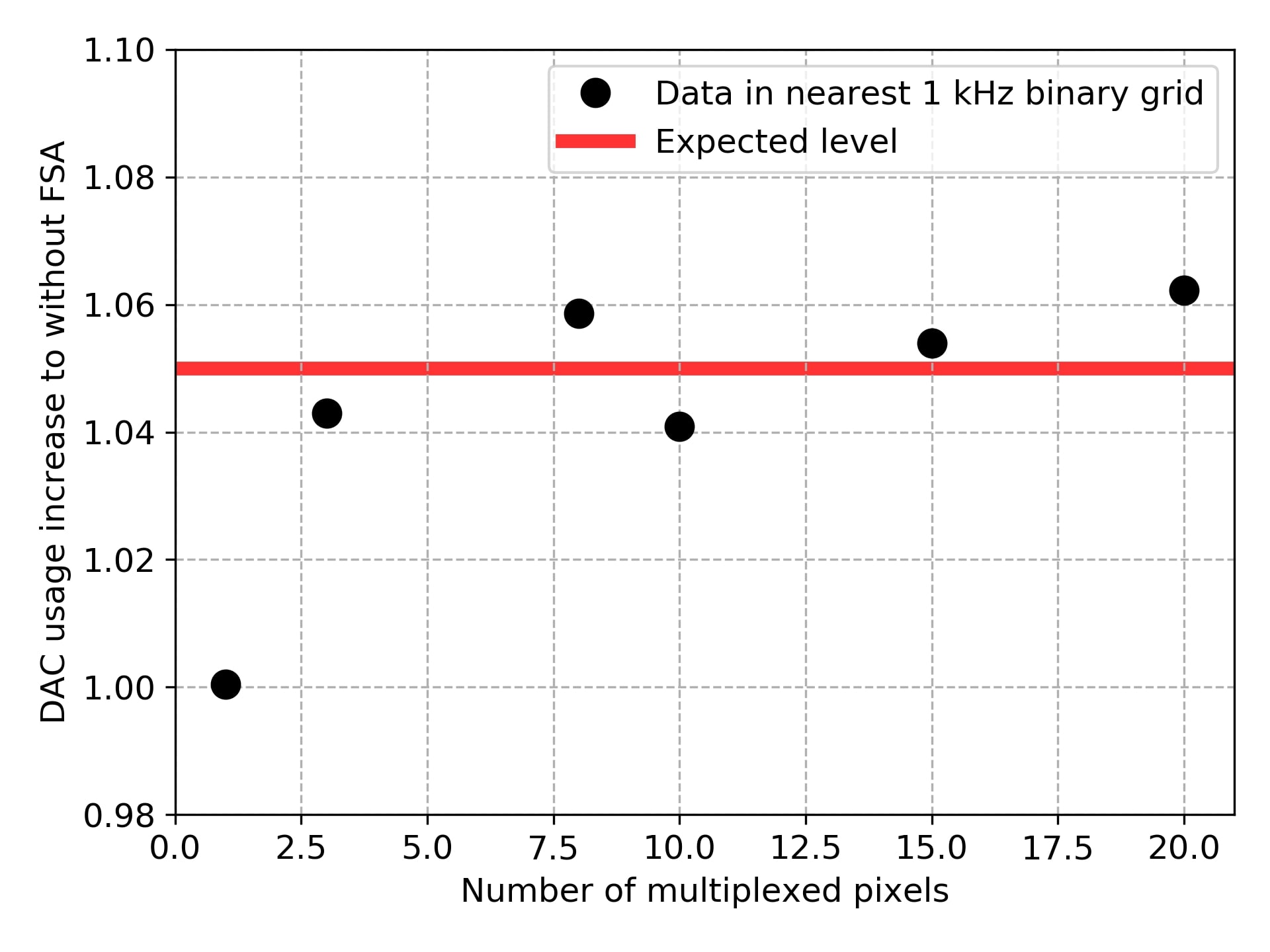}}
\caption{$Left$ $panel$: AC bias DAC voltage normalized to on-resonance value as a function of frequency shift, for 3 different bias points in the typical R/R$_{N}$ region. The resonance frequency used for reference is measured with the TES biased at 20\% of its transition (black points). The asymmetry of the parabola for the red and blue points is due to the slight change in resonance frequency at different bias points. $Right$ $panel$: voltage fed by the bias DAC with FSA active normalized to the on-resonance value as a function of active pixels. The normalised  DAC voltage for one pixel active is close to 1 since for that pixel the shift towards its grid frequency is almost zero.}\label{dac}
\end{figure*}

To assess the impact on the bias DAC output, we measure its raw RMS voltage and rescale it to the full-range peak-to-peak value. We first characterize the increment in voltage as a function of the frequency shift, as shown in Figure~\ref{dac}, left panel. The measurement is done for one pixel in three different bias points in the typical operation range. For lower TES resistance values, the $Q$-factor of the resonance is higher and so is the increase in the additional impedances when moving away from resonance (0 frequency shift). For this reason, the increment in DAC signal is larger for lower bias points.

Since for this setup the average shift to place the carriers on a 1 kHz grid is approximately 250 Hz, from this measurement we expect a DAC voltage increase of $\approx$ 5\%. To verify this, we measure the DAC voltage as a function of the number of multiplexed pixels, with bias frequencies shifted to fall on the 1~kHz grid, up to the representative number of 20 pixels. The result is reported in the right panel of Figure~\ref{dac}. We observe that the DAC voltage increase is compatible with the expected value of 5\% for R/R$_{N} \approx 0.2$ for an average shift of $\approx$250~Hz.

We conclude that the average frequency shift to place the carriers on grid allows to make a fair prediction of the increase in DAC voltage, which can be exploited to set a safety margin in the used dynamic range to avoid DAC saturation.

By design, the controller provides a signal amplification that compensates for the LC filter attenuation: however in this scheme also noise is passed through the feedback loop and amplified, thus increasing the effective DAC noise contribution to the spectral performance. 

To make a quantitative evaluation of the impact on the DAC noise contribution to the energy resolution, we refer to the estimated energy resolution budget for the FDM readout of X-ray TES microcalorimeters in Athena X-IFU \cite{roland_dac}, where the total energy resolution of one pixel is considered as the root sum square of several contributions, among which 2~eV come from the detector noise in set point and 0.23~eV from AC bias DAC noise.

Considering only these two contributions, the pixel energy resolution is affected as follows:
\begin{equation*}
\sqrt{ \Delta E_{TES}^{2} + \Delta E_{DAC}^{2} } = \sqrt{ (2\ \textrm{eV})^{2} + (0.23\ \textrm{eV})^{2} } \approx 2.01\ \textrm{eV}\ .
\end{equation*}

When FSA is active and frequency shift is applied, the DAC noise is amplified along with the signal. Calling $\varepsilon$ the increase in signal fed by the bias DAC, the estimated spectral performance then becomes
\begin{equation*}
\Delta E_{FSA} \equiv \sqrt{\Delta E_{TES}^{2} + (\varepsilon\cdot \Delta E_{DAC})^{2} }.
\end{equation*}

To avoid underestimations, since we are not taking into account other effects such as the increase of DAC non-linearity, we assume the worst possible case: all the pixels need a 700~Hz shift (approximately the maximum observed shift for a 1~kHz grid operation with this setup). Considering the blue curve from Figure~\ref{dac}, this means a voltage increase of roughly 40\%. With $\varepsilon = 1.40$ we then get $\Delta E_{FSA} \simeq 2.03$~eV.

Therefore, even making a substantial overestimation, we expect the increase in DAC noise to not contribute significantly to the root sum square spectral performance, in particular compared to the degradation observed due to (1) off-resonance operation without FSA \cite{hiroki2018} and (2) the presence of IMLN's within the thermal bandwidth, as shown in previous subsection.

One more observation is that, to perform these measurements, the system could handle stable FSA operation with more than 20 pixels active. This is a fundamental step forward with respect to past FSA implementations and a crucial point for practical FDM application on X-ray TES microcalorimeters.

\section{Bias circuit reproducibility}\label{iv_elec}

The basic functionality of FSA is to preserve the electrical bias circuit when shifting the carrier from the resonance frequency. In particular, the controller must react if the measured TES current is not in-phase with the bias voltage. Since even on resonance a phase angle can be present, due to unavoidable delays in the signal path, the phase of the carrier is tuned to compensate for such static offset. This is done with an automated routine that scans for the phase angle that minimizes the demodulated $Q-$signal with the TES in normal state. In this way, the controller correctly keeps the TES current in phase with the bias voltage when frequency shift is applied.

To verify the conservation of the bias circuit, we measure the TES IV curve with FSA under frequency shift and compare it with the original IV curve on-resonance. Early tests showed that the IV curves diverged for increasing frequency shifts. This was understood to be the effect of non-zero admittance of the other resonators in parallel with the measured pixel. On resonance, the TES impedance is much lower than the parallel impedance of other resonators and the effect is negligible. However, when applying a frequency shift the impedance of the LCR series gets higher and the contribution of the parallel impedance becomes significant.

The compensation for such effect is made in the firmware, by multiplying the bias signal for a scaling factor, essentially the parallel admittance converted to adimensional units. This scaling factor is given by numerical parameters depending on the setup (LC resonances, inductance value, SQUID coil ratio, bias and feedback transfer function, conversion factors of warm electronics) and is readily calculated for each pixel. In addition, we developed an iterative procedure that empirically optimizes such scaling factor, which we use to verify the goodness of the calculation. This approach allows us to correctly compensate for the leakage of current though the parallel resonators.

As previously stated, the electrical bandwidth at typical bias points for our pixels is of the order of 1~kHz. Within this boundary ($\pm$1~kHz shift) the controller should preserve the electrical circuit in an acceptable way. From single-pixel X-ray measurements with this array we know that the best performances are obtained by biasing the pixels approximately from 15\% to 25\% of TES transition. From the IV curves we verified that in such transition range, a change in current of the order of 1\% is negligible compared to the typical period of the $Q$-current oscillation due to weak-link effect. Since no significant impact on spectral performance is expected from such small effect for these devices, we set as a requirement that the controller should maintain the electrical circuit within less than 1\% deviation from reference. To verify this, we measure how the TES current deviates from its on-resonance value when frequency shift is applied. This is done by measuring the TES current (1) as a function of bias voltage (IV curve) with fixed frequency shift and (2) as a function of frequency shift (current scan) with fixed bias voltage. 

\begin{figure}
\includegraphics[width=8.5cm]{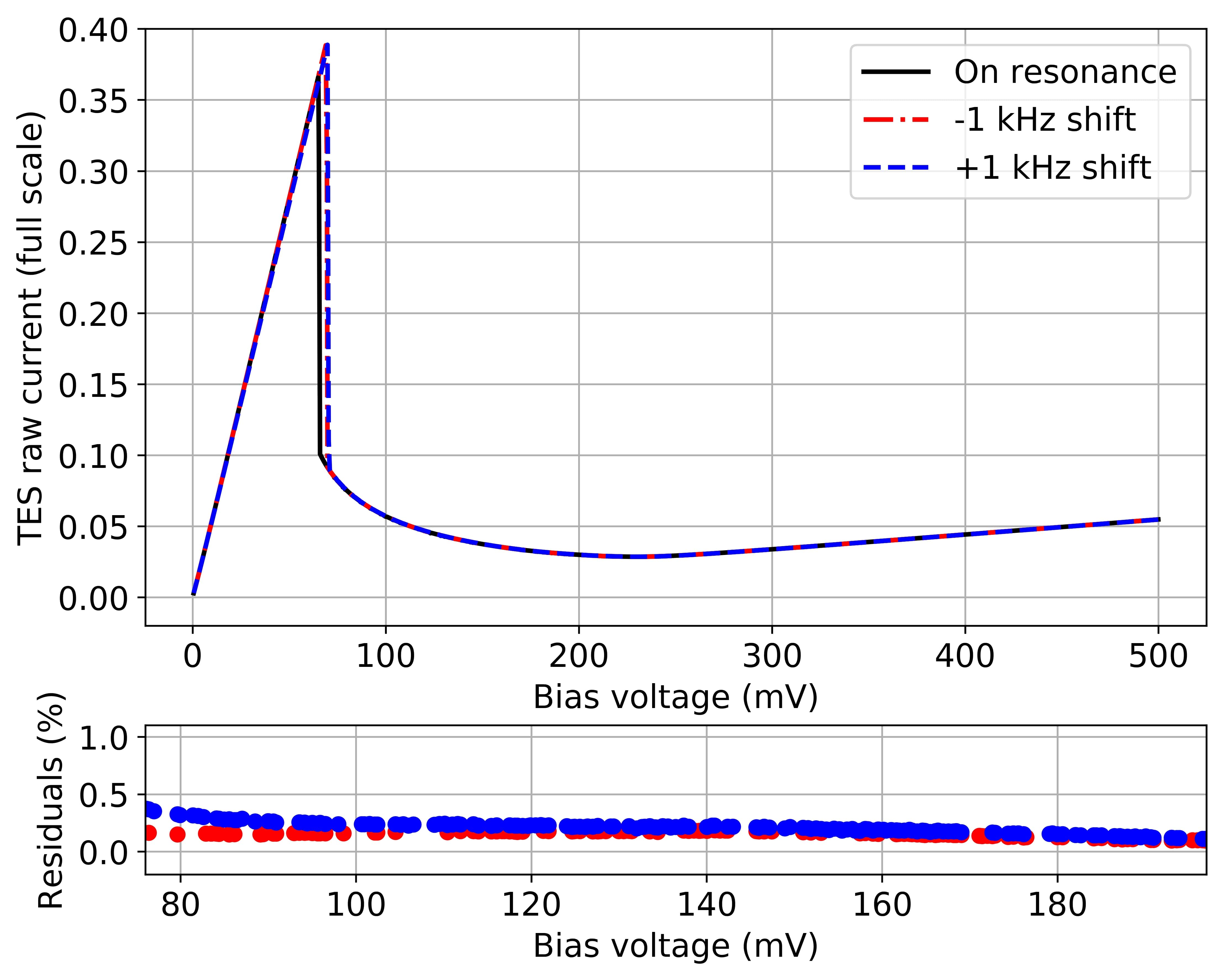}
\caption{Comparison of the IV curves on-resonance and shifted with FSA active. In the bottom panel, the residuals of the shifted curves with respect to the on-resonance reference, zoomed in the superconducting transition region are plotted scaled in percent. Some points, recorded in coincidence with an X-ray hit, are removed from the plot for sake of clarity.}\label{iv}
\end{figure}

\begin{figure}
\includegraphics[width=9cm]{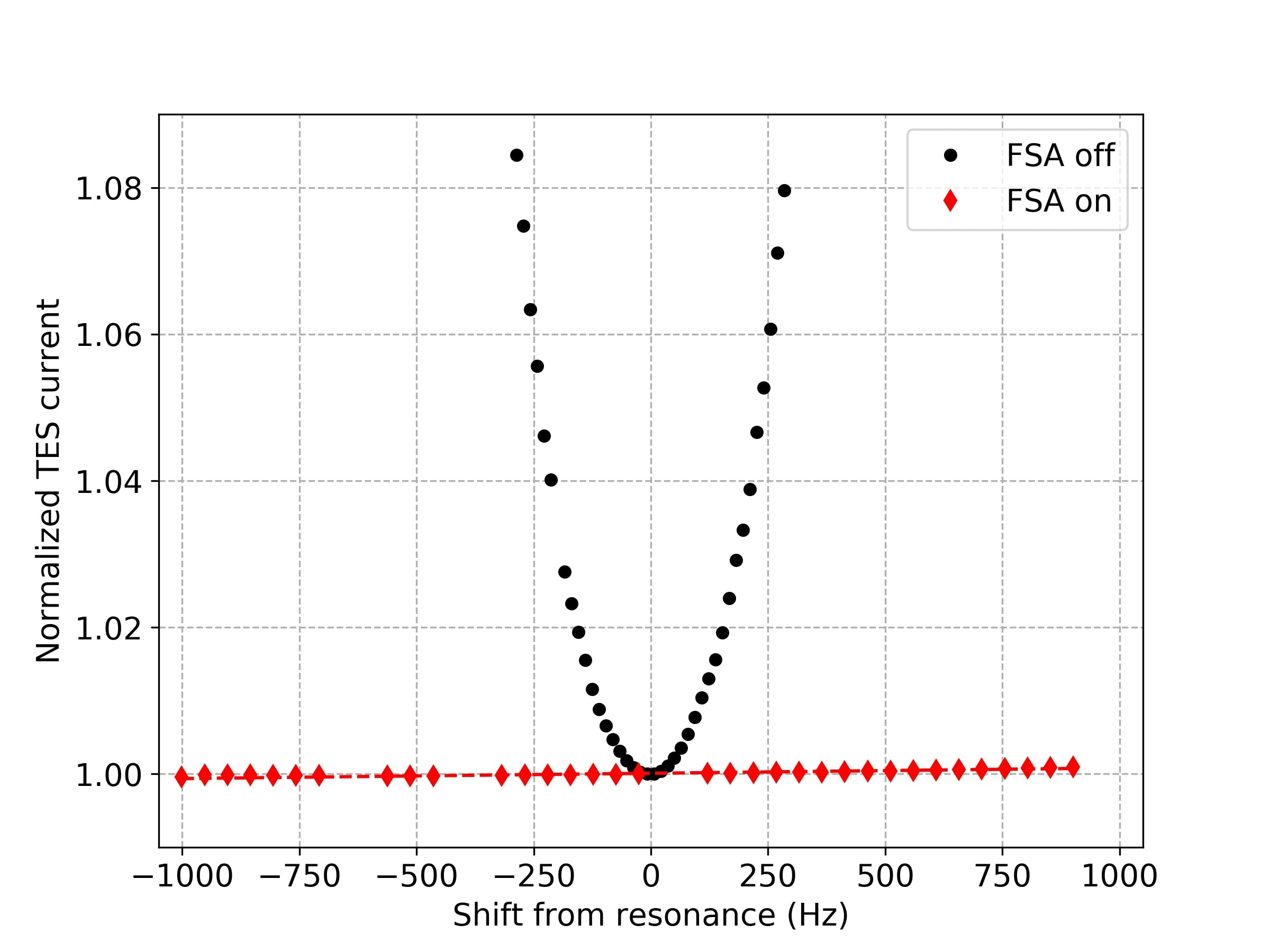}
\caption{TES current change on typical bias point with FSA active across $\pm$1 kHz frequency shift. Without FSA it is expected to have a similar parabolic shape as the DAC compensation voltage (Figure~\ref{dac}, left panel), since both come from the effect of the additional impedance due to off-resonance operation. Some points, recorded when an X-ray hit the detector, are not visible in the shown scale.}\label{iv_cur}
\end{figure}

We verified that, by properly compensating the parallel admittance effect, we are able to (1) reproduce the IV curve at shifted frequency and (2) keep the TES current constant at the typical working point within our boundary conditions ($\pm$1~kHz shift, $<$ 1\% deviation). In Figure~\ref{iv} the shifted IV curves and in Figure~\ref{iv_cur} the TES current scan are representatively reported for one pixel.

\section{X-ray spectral performance}\label{results}

As previously anticipated, the controller should be fast enough to follow the rapid decrease of the TES current when an X-ray photon hits the detector. The parameters to set are essentially three: the integrator gain, the crossover frequency $f_{PI}$ (defined in Section~\ref{requirements}) and the low-pass filters cut-off frequency $f_{LPF}$.

The procedure used to set these parameters is the following: to allow the controller to correctly compensate the TES current during an X-ray event, we increase the integrator gain until the TES response is stable and the X-ray pulse shows no signs of oscillation. We find that the integrator gain should be increased approximately in proportion with the TES bias frequency. In practice, we use 3 sequential values for 3 corresponding regimes, which we define as low- ($\approx$~1~MHz to 2~MHz), mid- ($\approx$~2~MHz to 3~MHz) and high-frequency ($\approx$~3~MHz to 5~MHz). To allow a sufficient integrator gain while maintaining the feedback stable, the proportional gain is tuned so that the crossover frequency $f_{PI}$ covers the electrical bandwidth of the pixel, with a factor $\approx$ 2-3 of margin. The cut-off frequency $f_{LPF}$ of the low-pass filter is then set to $\approx$ 2-3$\times f_{PI}$, value for which the out-of-band Q-feedback signal is sufficiently suppressed to prevent interference with 100~kHz neighbour channels in multi-pixel operation, but with enough margin to avoid attenuation of the in-band TES response. We find that stable operation for all the pixels can be achieved by using the same values of $f_{PI}$ and $f_{LPF}$.

Having fixed such parameters, we first perform X-ray measurements in single-pixel with 200~Hz frequency shift for several channels, confirming that the energy resolution does not degrade under FSA operation. In this way we confirm that the controller is able to compensate for fast changes in TES current, while maintaining stable operation. In Table~\ref{tablesp} the comparison between the typical single pixel performance and the measured performance with FSA is representatively reported for one pixel per each frequency regime.

\begin{table}[!h]
\begin{center}
    \begin{ruledtabular}
    \begin{tabular}{ cccc }
    \textbf{Ch nr.} & \textbf{Freq (MHz)} & \textbf{$\Delta$E$_{\textrm{SP}}$ (eV)} & \textbf{$\Delta$E$_{\textrm{SP}}$ FSA (eV)} \\ \hline
    1 & 1.1 & 2.57 $\pm$ 0.12 & 2.57 $\pm$ 0.17 \\ 
    17 & 2.9 & 2.76 $\pm$ 0.11 & 2.82 $\pm$ 0.17 \\
    25 & 4.2 & 2.75 $\pm$ 0.12 & 2.83 $\pm$ 0.20 \\
    \end{tabular}
    \end{ruledtabular}
\end{center}
\caption{Comparison with typical single-pixel performance (averaged over two different bias points) and measured single-pixel performance with FSA under a 200 Hz frequency shift.}\label{tablesp}
\end{table}

\begin{figure*}
\subfigure{\includegraphics[width=8.7cm]{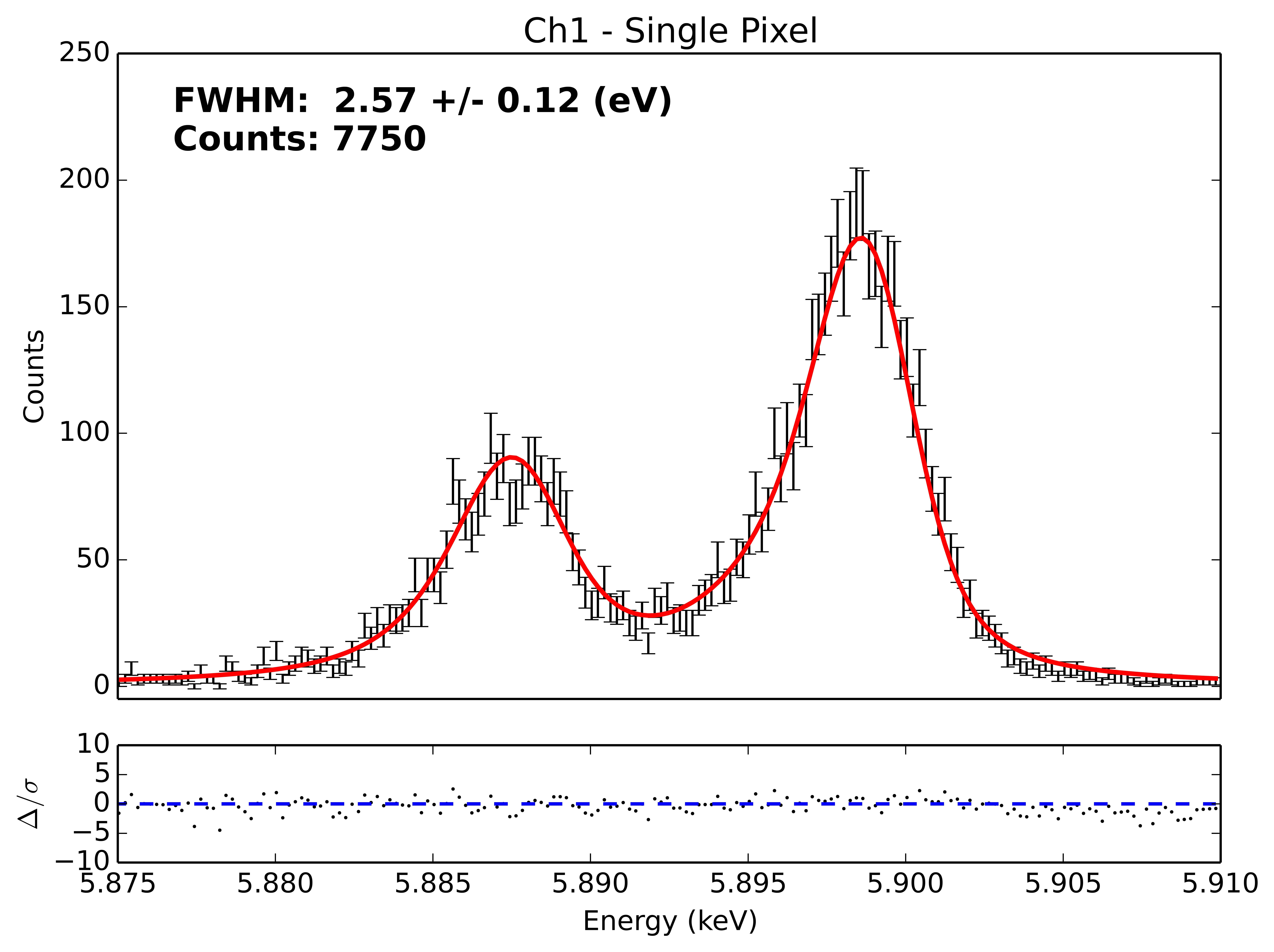}}
\subfigure{\includegraphics[width=8.7cm]{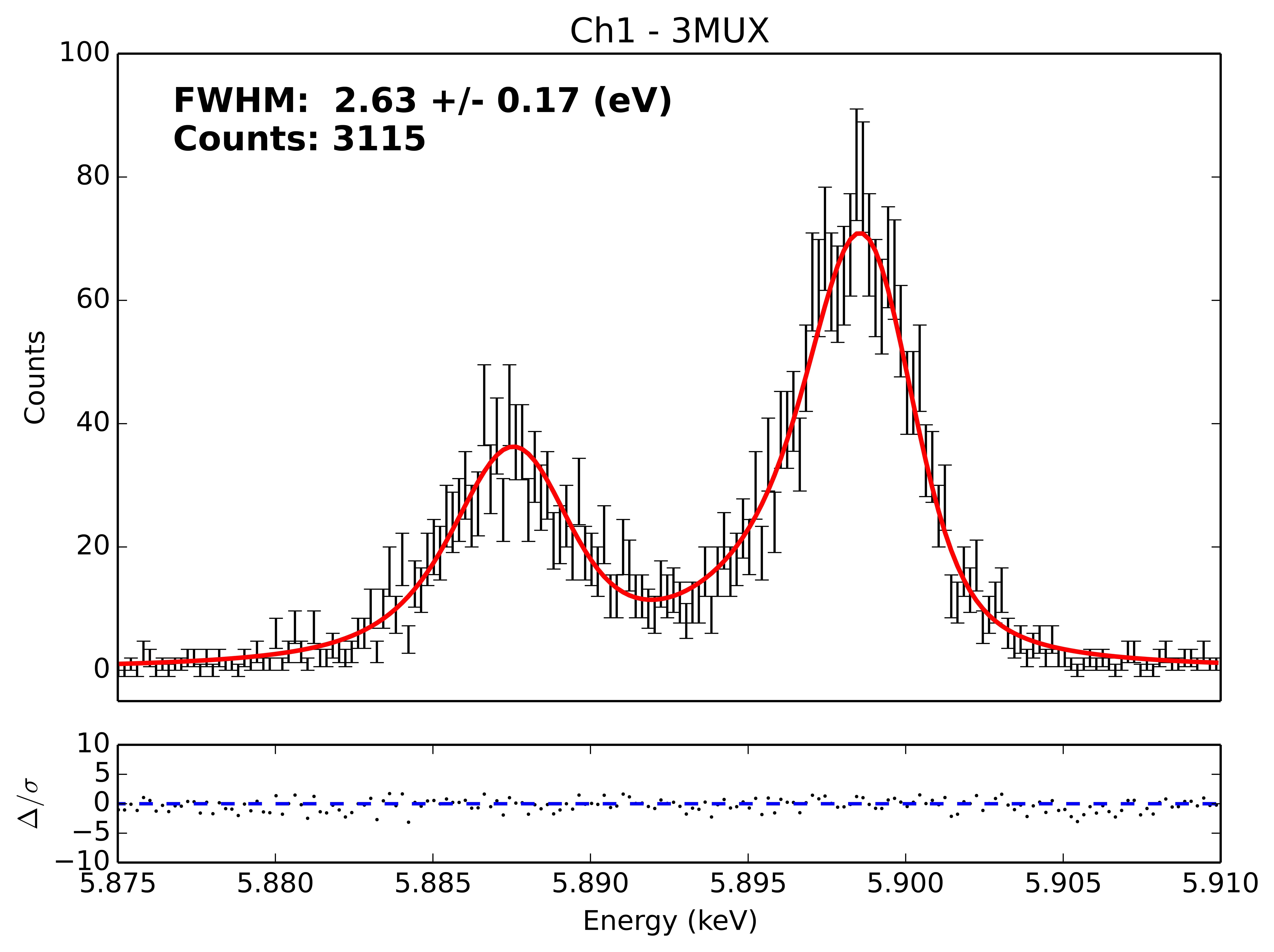}}\\
\subfigure{\includegraphics[width=8.7cm]{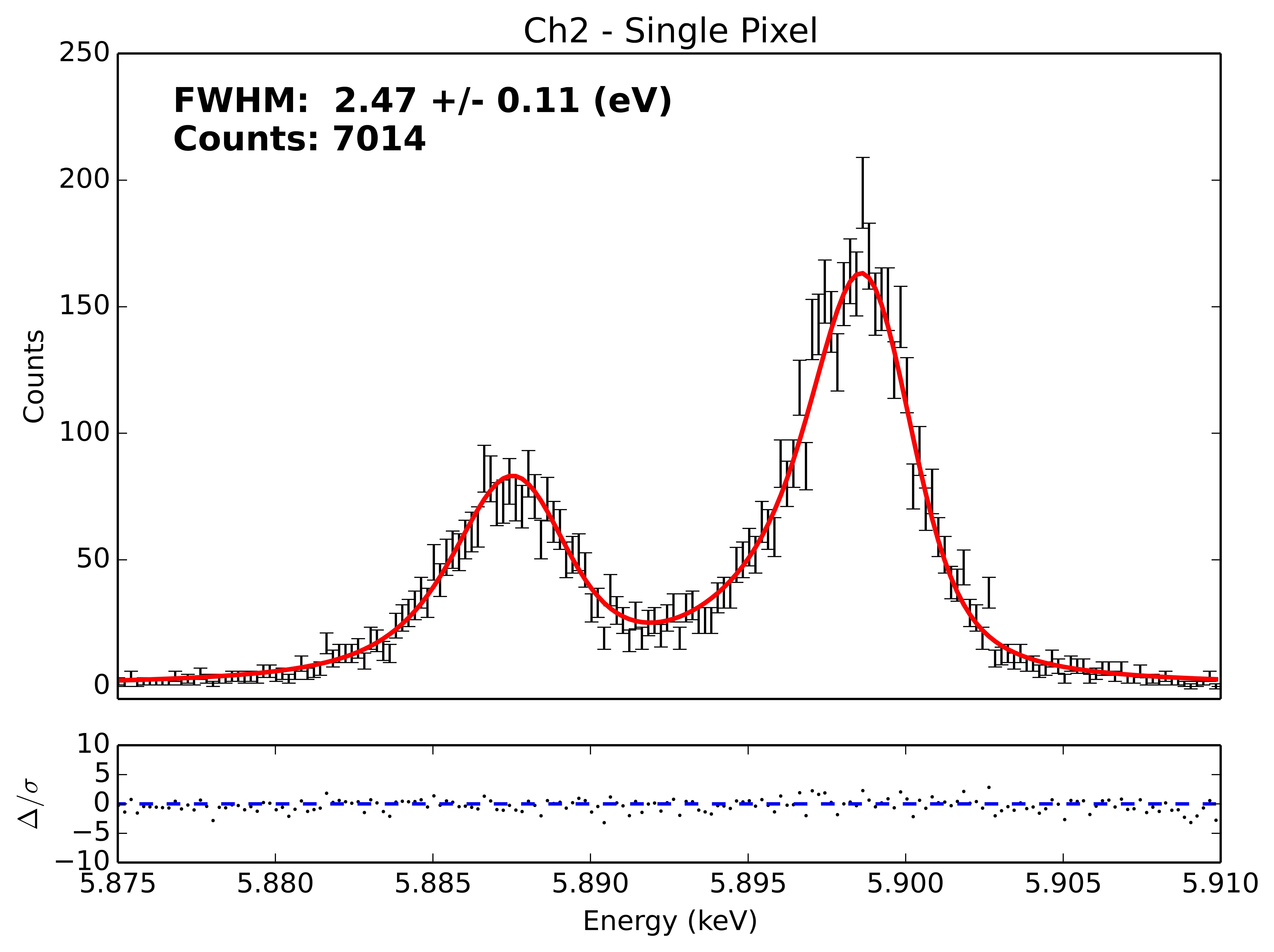}}
\subfigure{\includegraphics[width=8.7cm]{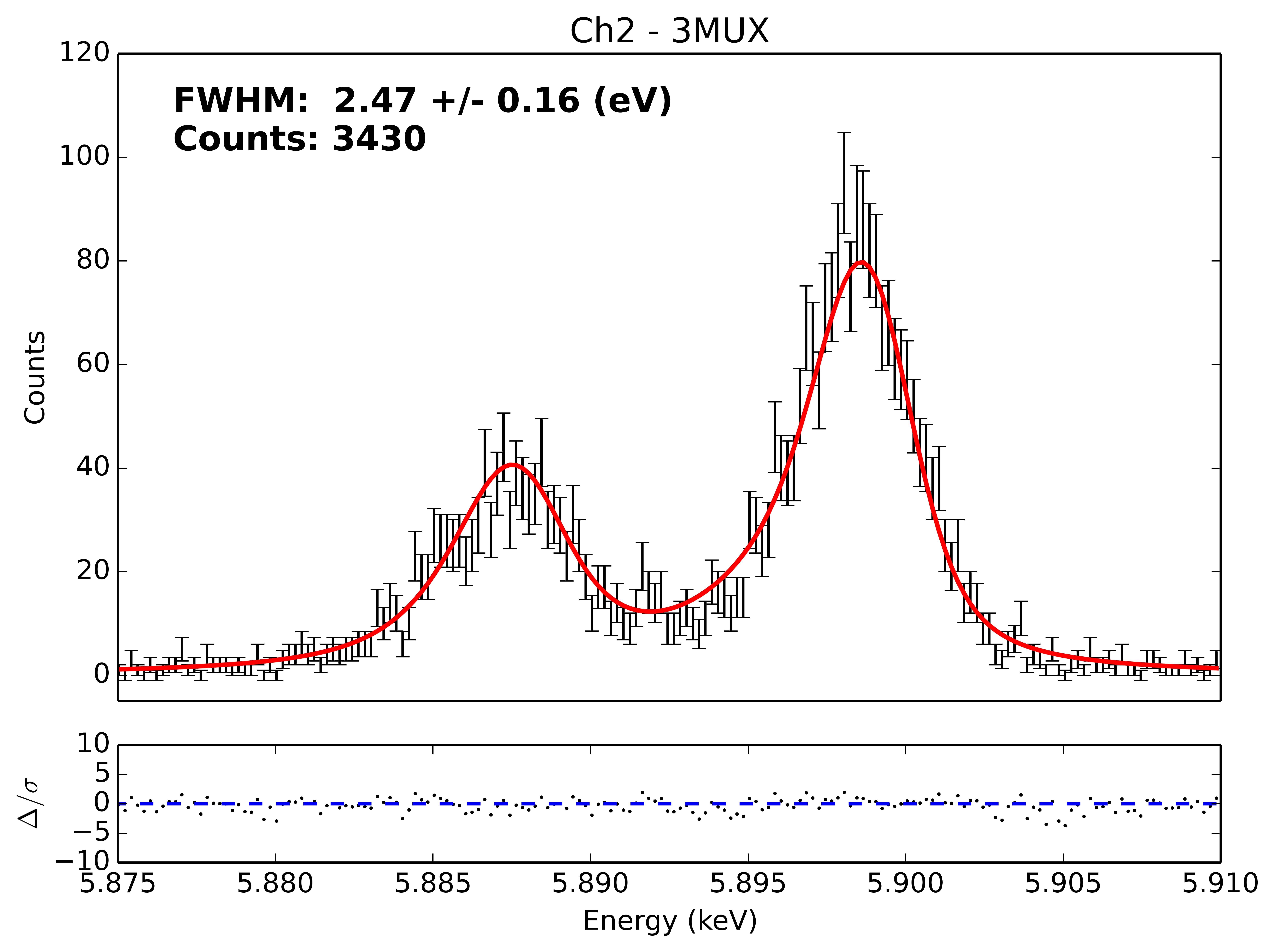}}\\
\subfigure{\includegraphics[width=8.7cm]{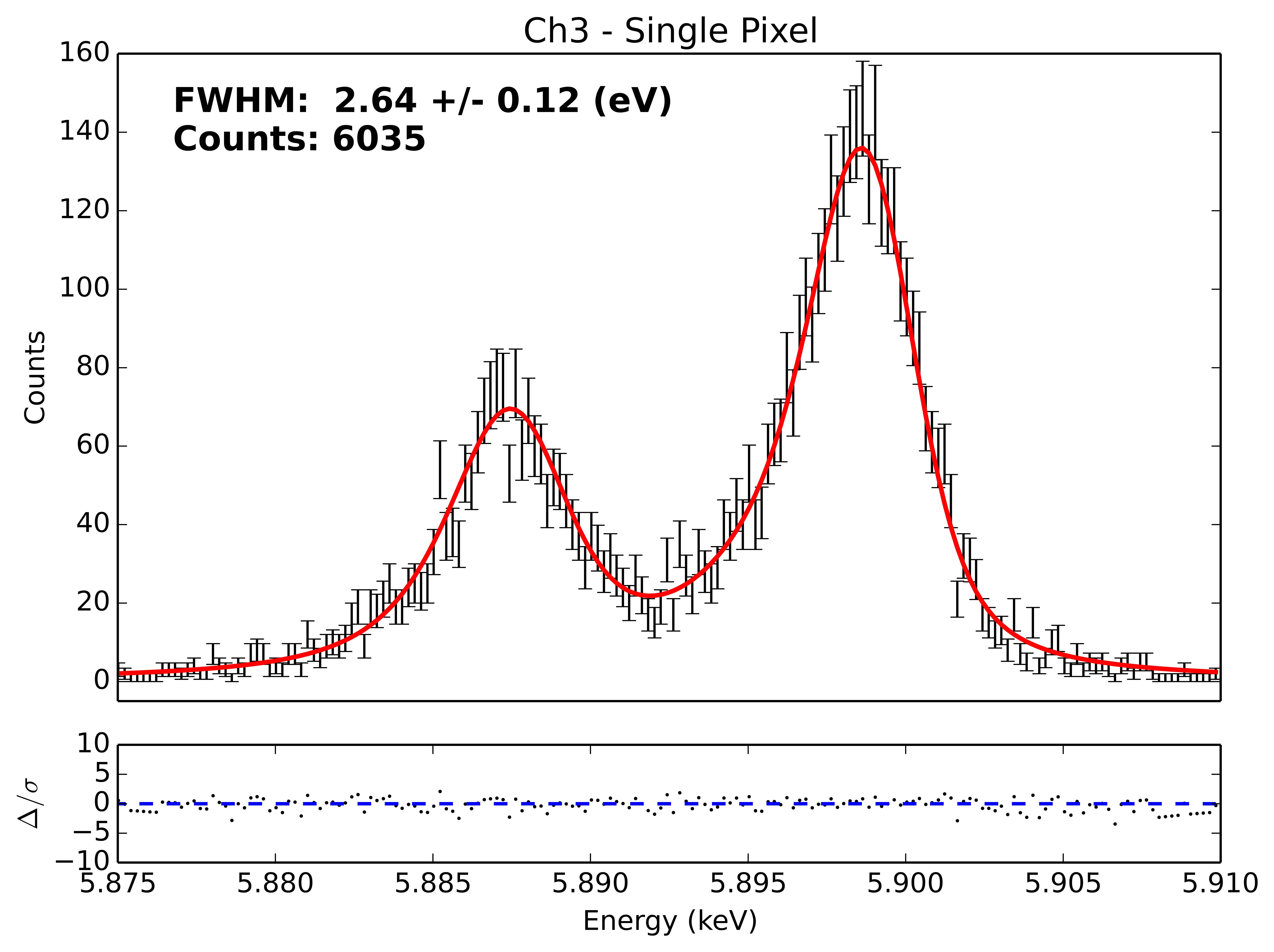}}
\subfigure{\includegraphics[width=8.7cm]{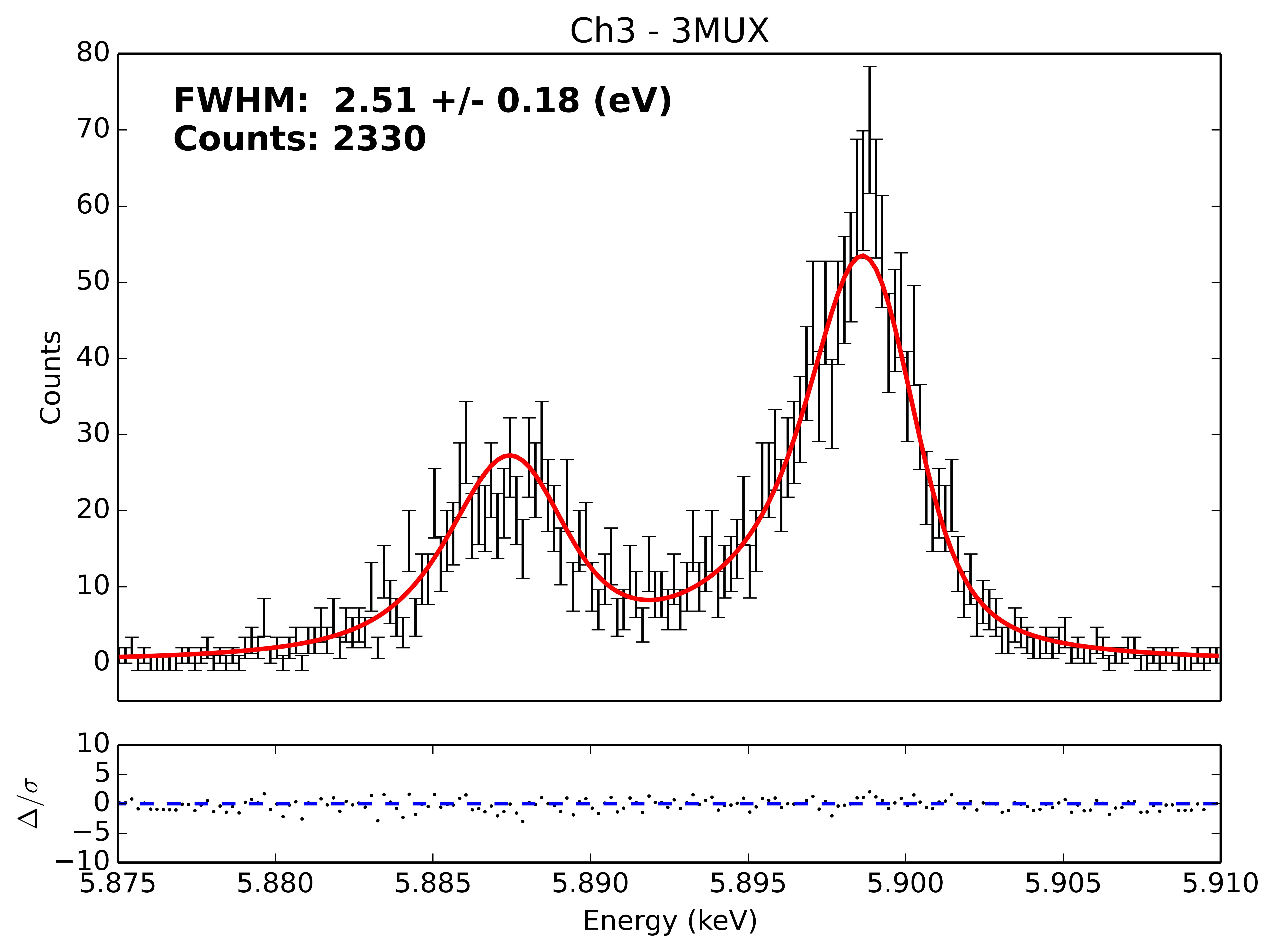}}\\
\caption{Measured spectral resolutions for Ch1-2-3 in single pixel, on resonance and without FSA (left panels), and in multi-pixel at 100 kHz separation, with FSA active and bias frequencies shifted on a 1~kHz grid (right panels).}\label{3mux}
\end{figure*}

We then perform multiplexing experiments to verify that the energy resolution is not degraded in multi-pixel configuration. We first use three neighbouring channels in the 100 kHz separation, namely Ch1, Ch2 and Ch3, whose single-pixel average performance is at level of $\approx$~2.5~eV. The bias frequencies are placed on a regular 1~kHz grid, resulting in shifts of approximately 150 Hz for Ch1 and Ch2 and 600 Hz for Ch3. Measuring the X-ray performance in multiplexing we find a summed energy resolution of 2.54~$\pm$~0.10~eV, consistent with the expected value from single-pixel measurements. The energy spectra for these channels in single-pixel and 3-pixel multiplexing are compared in Figure~\ref{3mux}.

Using the same FSA parameters, we repeat the measurement considering other pixels in the 100 kHz separation (namely Ch5, Ch6 and Ch7). In this case, the frequency shifts necessary for on-grid operation are $\approx$~130~Hz, 210~Hz and 420~Hz, respectively. We measure a summed average energy resolution of 2.55 $\pm$ 0.09 eV, which again is consistent with the expected value from single-pixel measurements. The results of such measurements are summarized in Table~\ref{tablemux}.

\begin{table}
\begin{center}
    \begin{ruledtabular}
    \begin{tabular}{ cccc}
    \textbf{Ch nr.} & \textbf{Freq (MHz)} & \textbf{$\Delta$E$_{\textrm{SP}}$ (eV)} & \textbf{$\Delta$E$_{\textrm{3MUX}}$ FSA (eV)} \\ \hline\
    1 & 1.1 & 2.57 $\pm$ 0.12 & 2.63 $\pm$ 0.17\\ 
    2  & 1.2 & 2.47 $\pm$ 0.11 & 2.47 $\pm$ 0.16 \\
    3  & 1.3 & 2.64 $\pm$ 0.12 & 2.51 $\pm$ 0.18 \\ \hline
    5 & 1.5 & 2.65 $\pm$ 0.10 & 2.53 $\pm$ 0.15\\ 
    6  & 1.6 & 2.73 $\pm$ 0.11 & 2.71 $\pm$ 0.14 \\
    7  & 1.7 & 2.27 $\pm$ 0.10 & 2.42 $\pm$ 0.15 \\
    \end{tabular}
    \end{ruledtabular}
\end{center}
\caption{Energy resolutions at 5.9 keV measured in single pixel (no FSA) and comparison with the individual performances obtained in 3-pixel multiplexing experiments with FSA, with 100 kHz neighbour channels and bias frequencies shifted on grid.}\label{tablemux}
\end{table}

These experiments allow us to verify that (1) FSA is stable in multi-pixel operation with 100 kHz separated pixels and (2) once the optimal values for $f_{PI}$ and $f_{LPF}$ are fixed, no further fine tuning is required for the different pixels and/or frequency shifts. With respect to the previous version of FSA, these are major points of improvement that consent to scale the applicability of FSA up to an entire readout chain.

\begin{figure*}
\includegraphics[width=10cm]{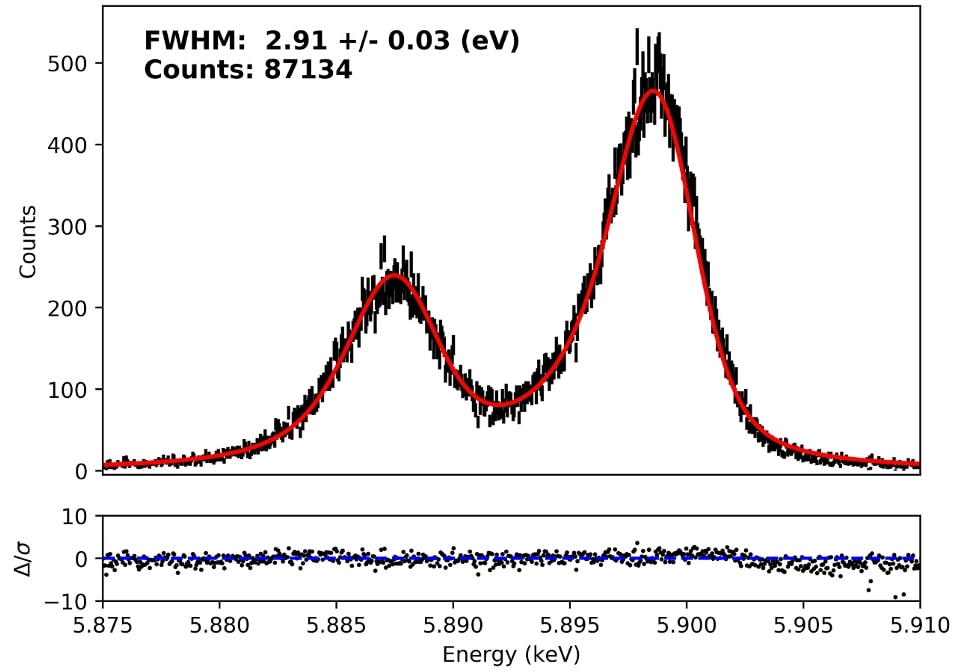}
\caption{Summed energy spectrum from 20 pixels measured in a 22-pixel multiplexing experiment (2 outliers excluded in the analysis) with FSA.}\label{histsum}
\end{figure*}

\begin{figure*}
\includegraphics[width=16cm]{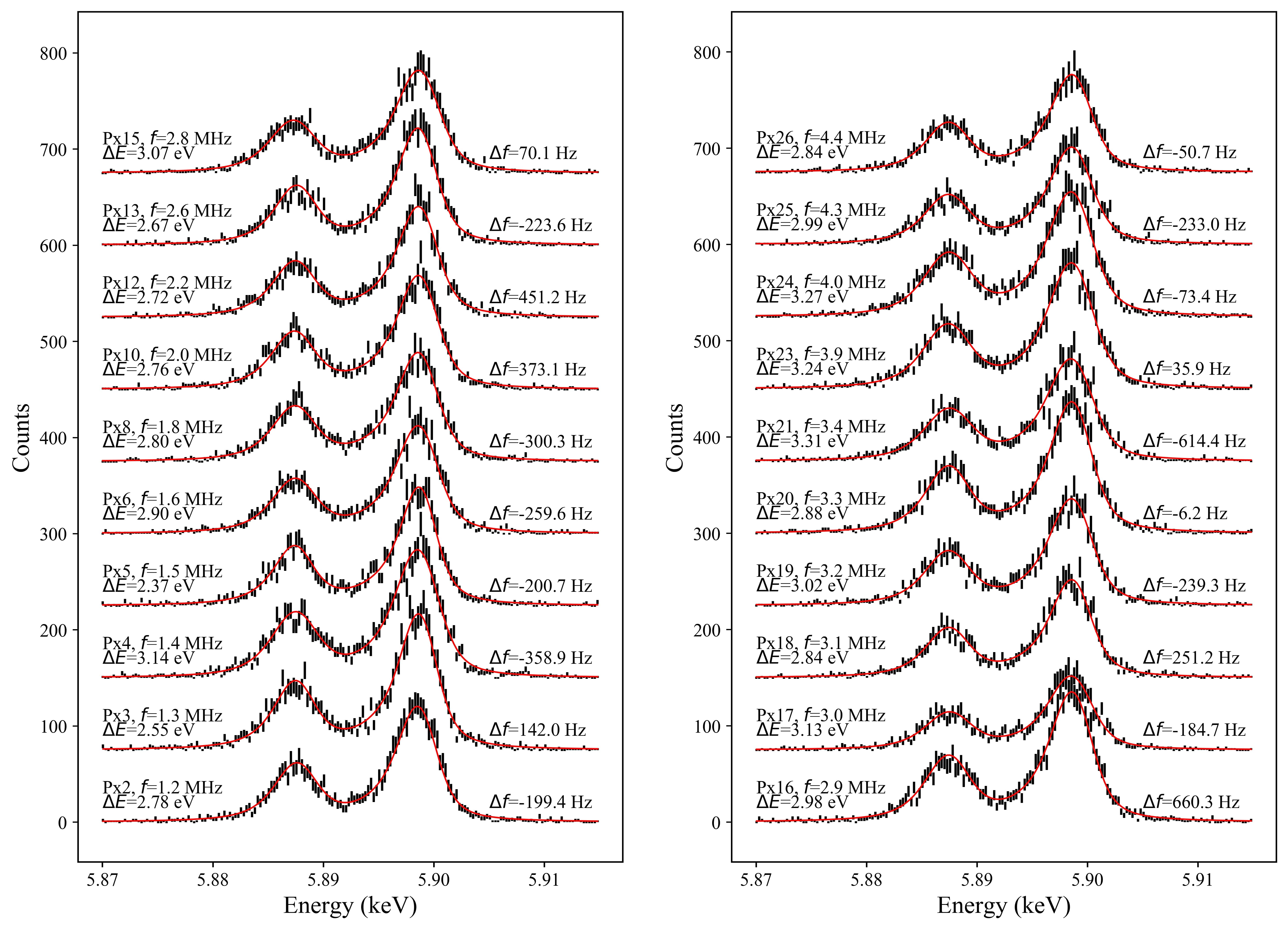}
\caption{Individual energy histograms of 20 pixels measured in a 22-pixel multiplexing experiment (2 outliers excluded in the analysis) with FSA. Bias frequency, energy resolution and shift from resonance to on-grid frequency are reported for each pixel in the bottom panel.}\label{hist20}
\end{figure*}

To show FDM results is outside the focus for this paper, nonetheless we mention that we verified that in $\geq$~20-pixel multiplexing we can obtain stable operation of the system and summed spectral resolution at a level better than 3~eV, where the typical average energy resolution measured in single pixel with this array is at a level of 2.7-2.8~eV, with one outlier at a level of 4.4~eV and two at a level of 3.5~eV. In Figure~\ref{histsum} and Figure~\ref{hist20} we report, respectively, the summed histogram and individual energy spectra of 20 pixels obtained in a 22-pixel multiplexing experiment (the 4.4~eV outlier not being active), with a measured summed energy resolution at 5.9~keV of 2.91$\pm$0.03~eV. The summed energy resolution degrades to 2.97$\pm$0.03~eV if the two 3.5~eV outliers are included in the analysis.

A further, more detailed discussion on FDM results, obtained with new SRON arrays\cite{martin2020}, will be reported separately in a paper currently under preparation.

\section{Summary}

We have developed a frequency shift algorithm to perform FDM readout of TES detectors in off-resonance operation. In this scheme, by shifting the carriers on a regular grid, line noises due to intermodulation distortion can be placed sufficiently far away from the sensitive region on the TES response, avoiding degradation in spectral performance. By applying the FSA up to 20 pixels in multiplexing, we evaluated that the increase in bias DAC output is acceptable in terms of DAC dynamic range and noise contribution to the energy resolution.

After finding and fixing the optimal FSA parameters for our experimental setup, we have verified that the on-resonance bias circuit can be reproduced with acceptable margin when the bias frequency is shifted.

Finally, we have measured the X-ray spectral performance in multi-pixel configuration in 100~kHz neighboring pixels with off-resonance, on-grid bias frequencies, observing no significant degradation from single-pixel energy resolutions. This is achieved without the necessity of fine-tuning the controller parameters for the different pixels and frequency shifts.

Having shown scalability towards and stability in multi-pixel operation, we consider the FSA mature enough to be implemented on the FDM readout of X-ray TES microcalorimeters, to finalize the demonstration of our FDM technology.

\section*{Acknowledgements}

SRON is financially supported by the Nederlandse Organisatie voor Wetenschappelijk Onderzoek.

The SRON R1a TES array used for the measurements reported in this paper was developed in the framework of the ESA/CTP grant ITT AO/1-7947/14/NL/BW.

\section*{Data availability}

The corresponding author makes available the data presented in this paper upon reasonable request.

\end{document}